%% file: main.tex
\documentclass[reprint,amsmath,amssymb,aps,prl,superscriptaddress, showpacs, eprint]{revtex4-2}

\input{preamble}

\usepackage{here}

\begin{document}
\title{Revealing quantum geometry of solids through vacuum fluctuations}
\author{Yugo Onishi}
\email{yugo0o24@mit.edu}
\affiliation{Department of Physics, Massachusetts Institute of Technology, Cambridge, MA 02139, USA}
\author{Liang Fu}
\affiliation{Department of Physics, Massachusetts Institute of Technology, Cambridge, MA 02139, USA}

\date{\today}

\begin{abstract}
	We show that quantum fluctuations of electromagnetic fields induce an extensive zero-point energy in insulators. The zero-point energy density is proportional to the quantum fluctuation of electric polarization in the many-body ground state, a fundamental quantum geometric property of solids known as the quantum weight.  
    For fields produced by a superconducting LC circuit, the zero-point energy results in a repulsive force between the circuit and the material
    and an attractive force on the capacitor plates, which we call quantum geometric force. The force on the capacitor plates is proportional to the quantum weight, inversely proportional to the plate area, and independent of the plate distance. The proposed effects provide direct experimental access to the many-body quantum geometry and reveal a new quantum geometric effect in macroscopic solids induced by vacuum fluctuation.
\end{abstract}

\maketitle

\textit{Introduction. ---}
A profound consequence of quantum field theory is that the vacuum is not empty. It is filled with fluctuating electromagnetic fields, leading to observable consequences. When coupled to these fluctuating fields, the atomic spectra of an atom experience a shift---called Lamb shift~\cite{lamb1947,bethe1947,welton1948}, and an attractive force between the atoms emerges---known as van der Waals force. These fluctuation-induced spectra shift and force are ubiquitous consequences of quantum electrodynamics (QED). 

In this century, developments in microwave cavities offer exciting opportunities to study the interaction between discrete electromagnetic modes and artificial atoms with an unprecedented tunability~\cite{haroche2006,blais2021}. The analogy of Lamb shift was observed in a superconducting qubit---an effectively two-level system---coupled to virtual microwave photons in a superconducting resonator~\cite{fragner2008,ao2023}. 

In this work, we study the effect of fluctuating electromagnetic fields on an insulator containing $\sim 10^{23}$ electrons, and show that the coupling to virtual low-energy photons results in an {\it increase} of the ground state energy of insulating states. This leads to predictions that can be experimentally tested: (1) a {\it repulsive} force between a dielectric and a microwave resonator that are {\it inductively} coupled, and (2) an attractive force on the capacitor plates in the microwave circuit. 

The energy shift and the force between a dielectric and a circuit should be contrasted with the Lamb shift~\cite{lamb1947,bethe1947,welton1948} and Casimir effect~\cite{casimir1948,casimir1948attraction}, which also originate from the vacuum fluctuations. The Lamb shift of the ground state is negative and the Casimir force is typically attractive, whereas we find a positive energy shift and a repulsive force between the solid and the resonator. Instead, the increase of the ground state energy and the resulting repulsive force can be understood as an analog of the repulsive Casimir or van der Waals force due to the diamagnetic processes~\cite{feinbergGeneralTheoryVan1970,boyerVanWaalsForces1974,buhmannGroundstateVanWaals2005,hoyeRepulsiveCasimirForce2018,kenneth2002,geyer2010,inui2011}.

Furthermore, this process enhances charge fluctuations of the capacitor of the circuit, leading to an attractive force between the capacitor plates. The attractive force is governed by a ground-state property of the insulator, scales as $1/S$ with the plate area $S$, and is independent of the plate distance $d$. These features distinguish it from the conventional Casimir force between the capacitor plates, which is insensitive to the insulator and typically scales as $S/d^4$~\cite{casimir1948attraction}.

Interestingly, the energy shift of an insulator and the resulting forces due to virtual microwave photons are directly determined by the quantum geometry of the many-body ground state. Since the energy scale of microwave photons is several orders of magnitude smaller than the electronic energy scale in solids, the electronic system is ``fast'' compared to the ``slow'' microwave fields. As a result, the electronic system remains in its instantaneous ground state, but acquires an additional zero-point energy that is extensive in its volume and proportional to the many-body quantum metric. Therefore, we shall call the predicted forces \textit{quantum geometric forces}. Our findings show that superconducting circuits are a powerful new tool for uncovering a fundamental ground state property of quantum materials---the many-body quantum metric---through the quantum geometric force.

\begin{figure}[htbp]
	\centering
	\includegraphics[width=0.9\columnwidth]{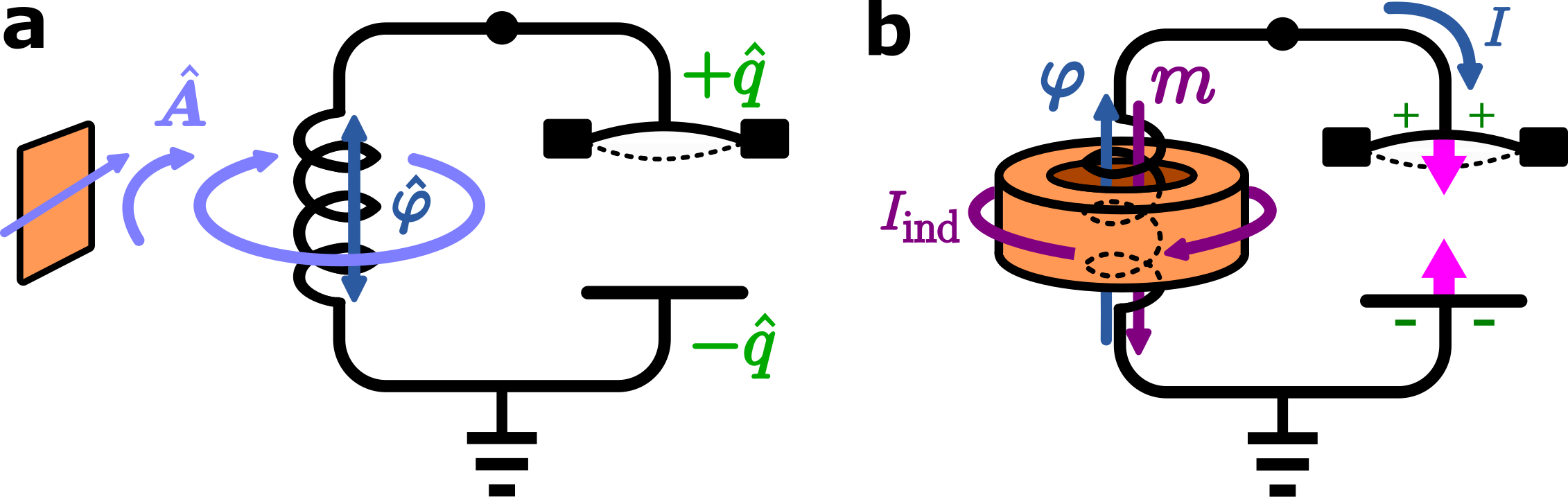}
	\caption{(a) Interaction between a superconducting LC circuit and a quantum material. The magnetic flux $\hat{\varphi}$ behaves as a magnetic dipole moment and creates the vector potential $\vec{A}$ on the electronic system. (b) The diamagnetic process due to the quantum fluctuation in the superconducting circuit. The flux $\varphi$ in the inductor induces the current $I_{\rm ind}$ in the electronic system, which results in a magnetic moment $m$ that is antiparallel to the flux $\varphi$, and thus increases the energy. The current $I$ accumulates the charge on the capacitor, resulting in an attractive force between the capacitor plates.}
	\label{fig:interaction}
\end{figure}

\textit{Theory. ---} We consider a general electronic system with a finite energy gap $\Delta$ between the ground state $\ket{\Psi_0}$ and the first excited state. Our purpose is to study the effect of the quantum fluctuation of electromagnetic fields on this system in the ground state. In particular, we consider the coupling between the electronic system and electromagnetic fields generated by a superconducting LC circuit, as illustrated in Fig.~\ref{fig:interaction}(a). 
The total system is described by the following Hamiltonian:
\begin{align}
H_{\rm tot} &= H[\hat{\vec{A}}(\vec{r}), \hat{\phi}(\vec{r})] %
+ H_p, \label{eq:tot_H1}
\end{align}
Here, $H[\hat{\vec{A}}(\vec{r}), \hat{\phi}(\vec{r})]$ describes the electronic system in the presence of the vector potential $\hat{\vec{A}}(\vec{r})$ and the scalar potential $\hat{\phi}(\vec{r})$, and can be either continuum or lattice (tight-binding) Hamiltonian.
$H_p$ is the dissipationless Hamiltonian for a superconducting LC circuit~\cite{girvin2014}:
\begin{align}
	H_p = \frac{\hat{\varphi}^2}{2L_p} + \frac{\hat{q}^2}{2C_p},
\end{align}
where $\hat{q}$ denotes the charge accumulated on the capacitor of capacitance $C_p$ and $\hat{\varphi}$ denotes the magnetic flux through the inductor of inductance $L_p$, which is related to the current $\hat{I}$ by $\hat{\varphi}=L_p\hat{I} + \mathrm{const}$.
They obey the canonical commutation relation $[\hat{\varphi}, \hat{q}] = i\hbar$. Regarding $\hat{\varphi}$ and $\hat{q}$ as the canonical coordinate and momentum, respectively, the Hamiltonian $H_p$ describes a quantum harmonic oscillator with resonance frequency $\omega_p = 1/\sqrt{L_pC_p}$. Thus, even in its ground state, the circuit exhibits zero-point fluctuations of charge, flux, and hence current.

The fluctuating charge and current in the circuit create scalar and vector potentials that act on the electronic system. %
From the perspective of electromagnetism, 
the capacitor acts as an electric dipole moment $P\propto \hat{q}/C_p$ and the inductor acts as a magnetic dipole moment $M\propto \hat{\varphi}$. 
Accordingly, we can generally write the potentials created by the LC circuit as
\begin{subequations}
\begin{align}
	\hat{\phi}(\vec{r}) &= g_C(\vec{r}) \frac{\hat{q}}{C_p}, \label{eq:gC_coupling} \\
	\hat{\vec{A}}(\vec{r}) &= \vec{g}(\vec{r}) \hat{\varphi},
\end{align}
\end{subequations}
where $g_C(\vec{r})$ and $\vec{g}(\vec{r})$ encode the spatial profiles of the electric and magnetic couplings, respectively, and are determined by the circuit geometry and by the distance to the capacitor and inductor.
Note that we chose the gauge such that $\ev{\hat{\vec{A}}}=0$ for $\ev{\hat{\varphi}}=0$. 

For simplicity, we assume that the spatial variation of the fields $\hat{\phi}$ and $\hat{\vec{A}}$ is negligible. This assumption is justified when the electronic system is much smaller than the LC circuit so that the potentials $\hat{\phi}(\vec{r})$ and $\hat{\vec{A}}(\vec{r})$ are approximately uniform within the system. 
In this limit, $g_C$ and $\vec{g}$ correspond to the mutual capacitance $C_m\sim g_C/C_p$ and the mutual inductance $M\sim l L_p g$, where $l$ is the linear size of the electronic system. We note that, when the electromagnetic fields vary slowly on the scale of the electronic wavelength, their effect is expected to be qualitatively the same as that of uniform fields. 

A constant scalar potential $\hat{\phi}$ has no effect on a charge-neutral system. By contrast, a spatially uniform vector potential is physically meaningful if it fluctuates. Indeed, its quantum fluctuation is gauge-invariant and directly related to the fluctuation of the electric field $\hat{E}_i$
as $\expval{\Delta \hat{E}_i\Delta\hat{E}_j}\propto \omega_p^2 \expval{\Delta\hat{A}_i\Delta\hat{A}_j}$, with $\Delta \hat{A}_i\equiv\hat{A}_i-\expval{A_i}$ and similar for $\Delta \hat{E}_i$.  
Thus, even a uniform vector potential can represent a physical electric field fluctuation, leading to observable consequences as shown below.

To analyze the Hamiltonian~\eqref{eq:tot_H1}, we employ the adiabatic approximation. Since energy gap of an electronic system is generally much larger than the energy scale of the circuit $\omega_p\lesssim \SI{10}{GHz}\sim\SI{10}{\micro\electronvolt}$, the electronic system is ``fast'' compared to the ``slow'' degree of freedom of the LC circuit, and therefore remains in the ground state corresponding to the instantaneous value of $\varphi$.
Under the adiabatic condition $\Delta \gg \hbar\omega_p$, we can obtain the effective Hamiltonian by performing a unitary transformation $U(\vec{A})$ that diagonalizes $H(\vec{A})$: 
\begin{align}
	&U^{\dagger}(\vec{g}\hat{\varphi})H_{\rm tot}U(\vec{g}\hat{\varphi}) = E(\vec{g}\hat{\varphi}) + \frac{\hat{\varphi}^2}{2L_p} + \frac{(q - \hbar\vec{g}\vdot\vec{a})^2}{2C_p}
\end{align}
where $E(\vec{A})$ is the ground state energy of $H(\vec{A})$, and $\vec{a} = i U^{\dagger} \nabla_{\vec{A}} U$ is the (non-Abelian) Berry connection associated with the ground state of $H(\vec{A})$. Then, by projecting onto the ground state of the electronic system,
we obtain the effective Hamiltonian as
\begin{subequations}
\begin{align}
	&H_{\rm eff} = E_0(\vec{g}\hat{\varphi}) + \frac{\hat{\varphi}^2}{2L_p} + \frac{(\hat{q} - q_0)^2}{2C_p} + E_{\rm QG} \label{eq:Heff_general} \\
	&E_{\rm QG} = \frac{e^2}{2 C_p}\vec{g} G \vec{g} \label{eq:E_QG}
\end{align}
\end{subequations}
where $E_0(\vec{A})$ is the ground state energy of $H(\vec{A})$, $q_0 = \hbar\vec{g}\vdot \vec{a}_{00}$ with $\vec{a}_{00}=\mel{\Psi_0(\vec{A})}{i\nabla_{\vec{A}}}{\Psi_0(\vec{A})}$ is the Berry connection associated with the ground state, and the tensor $G_{ij}$ is the quantum metric~\cite{wilczek1989,berry1997,aharonov1992,ohnishi2007} of the many-body ground state as a function of the vector potential: 
\begin{align}
    G_{ij} &= \qty(\frac{\hbar}{e})^2\Re\qty[\mel{\pdv{\Psi_0}{A_i}}{(1-P_0)}{\pdv{\Psi_0}{A_j}}]. \label{eq:G_def}
\end{align}
Here $P_0=\ketbra{\Psi_0}{\Psi_0}$ is the projection operator onto the ground state. Note that $G_{ij}\propto \hbar^2$ vanishes in the limit $\hbar \rightarrow 0$. For the periodic boundary condition, the expression~\eqref{eq:G_def} is equivalent to the quantum metric defined with the twisted boundary condition~\cite{niu1985}, as detailed in the Supplemental Materials (SM). 
 
$E_{\rm QG}$ in Eq.~\eqref{eq:Heff_general} emerges as the quantum correction to the adiabatic approximation. When $\vec{A}$ is treated as classical and slowly varying, the electronic ground state adiabatically follows the changing vector potential $\vec{A}$. Once the vector potential $\hat{\vec{A}}=\vec{g}\hat{\varphi}$ fluctuates quantum mechanically and couples to electrons in solids, the true ground state of the whole system is an entangled state of electrons and photons. Correspondingly, the ground state energy acquires a positive shift $E_{\rm QG}$ that is proportional to the quantum metric of the electronic system. 

Importantly, $E_{\rm QG}$ is finite even when the expectation value of electromagnetic fields vanishes. Rather, it relies on quantum fluctuation, $\Delta \hat{\vec{A}}$ and $\Delta\hat{\vec{E}}$. Only in the limit $C_p \rightarrow \infty$ where the flux $\varphi$ and therefore $\vec{A}$ becomes classical without fluctuations, the zero-point energy $E_{\rm QG} \propto 1/C_p$ vanishes. This should be contrasted to the standard electromagnetic properties of materials such as optical conductivity, where the applied electromagnetic fields are classical.

For systems with open boundary, the ground state under uniform $\vec{A}$ is related to the one under $\vec{A}=0$ through the gauge transformation $\ket{\Psi_0(\vec{A})} = e^{ie\vec{A}\cdot\vec{R}/\hbar}\ket{\Psi_0(0)}$, where $\vec{R} = \sum_n \vec{r}_n$ is the dipole moment operator of the electrons in the system. Using this relation, we can show that the many-body quantum metric is independent of $\vec{A}$ and given by
\begin{align}
	G_{ij} &= \expval{(R_{i} - \expval{R_{i}})(R_{j} - \expval{R_j})}. \label{eq:quantum_metric_open}
\end{align}
Therefore, $E_{\rm QG}$  can be expressed as 
\begin{align}
    E_{\rm QG} &=  \frac{\expval{(\vec{g}\cdot\Delta\vec{P})^2}}{2C_p} \label{eq:quantum_weight_open}
\end{align}
with $\Delta \vec{P} \equiv \vec{P} - \expval{\vec{P}}$ and the total polarization $\vec{P}\equiv e\vec{R}$.

Eqs.~\eqref{eq:Heff_general}, \eqref{eq:E_QG}, and \eqref{eq:quantum_weight_open} are the first main results of this work, establishing the effect of quantum vacuum fluctuation in a LC circuit on an electronic system.  
In the case of a single atom or molecule, $\vec{P}$ in Eq.~\eqref{eq:quantum_weight_open} is simply the electric dipole moment, and  %
$E_{\rm QG}$ reduces to the ground state expectation value of the dipole self-energy discussed in nonrelativistic QED~\cite{cohen-tannoudji2008,rokaj2018a}. For a macroscopic system, $G$ and therefore the zero-point energy $E_{\rm QG}$ scale with its volume, as we show below.

The expression~\eqref{eq:quantum_weight_open} clarifies the physical picture for zero-point energy $E_{\rm QG}$. Polarization fluctuation of the electronic system (an atom or a solid) induces a fluctuating current in the circuit through the inductive coupling, which enhances quantum charge fluctuation on the capacitor and increases the energy stored in the capacitor. 

We emphasize that the positive zero-point energy $E_{\rm QG}$  is a consequence of the inductive coupling (a current in the circuit induces a current in the electronic system) and is closely related to diamagnetism.
To see this, let us consider an electronic system forming a cylinder around the inductor, as shown in Fig.~\ref{fig:interaction}(b). In this case, the current in the electronic system along the circumferential direction can be viewed as a magnetic moment $m$, and the inductive coupling is equivalent to the coupling between the magnetic moments of the electronic system and the circuit. 
When the capacitance $C_p$ is finite, the current in the circuit fluctuates, producing a fluctuating flux $\varphi$ in the inductor. Then the electronic system responds \textit{diamagnetically} to oppose the change in the flux (Lenz's law), %
leading to an \textit{increase} in the energy of the system. This is the origin of the positivity of $E_{\rm QG}$.

Since $E_{\rm QG}$ is always positive, it contributes to a \textit{repulsive} force between the electronic system and the superconducting circuit. Since the coupling $\vec{g}$ is larger at a shorter distance between the electronic system and the inductor, the energy $E_{\rm QG}$ increases as they get closer, resulting in a repulsive force. 
Because this repulsive force is directly related to the quantum metric of the electronic insulator, we call it quantum geometric force.

The positive energy shift $E_{\rm QG}$ and the repulsive force should be contrasted with the Lamb shift of an atom~\cite{lamb1947,bethe1947,welton1948} and the van der Waals or Casimir force~\cite{casimir1948,casimir1948attraction,buhmann2012a}, where the quantum fluctuation of electromagnetic fields typically results in \textit{negative} ground state energy correction and an \textit{attractive} force between two atoms or macroscopic bodies. Rather, the repulsive quantum geometric force is an analog of the repulsive Casimir force between dielectric and magnetic systems ~\cite{feinbergGeneralTheoryVan1970,boyerVanWaalsForces1974,buhmannGroundstateVanWaals2005,hoyeRepulsiveCasimirForce2018,kenneth2002,geyer2010,inui2011}. 

We emphasize that $E_{\rm QG}$ and the quantum geometric force are determined by the ground state property of the electronic system, $G$. This is in contrast to the Casimir force, which typically involves infinitely many electromagnetic field modes spanning the whole frequency regime and thus depends on the excitation spectrum in general. For example, the conventional Casimir force is determined by the dielectric response function~\cite{klimchitskaya2009}, and cannot be determined only by the ground state properties.  
We note that an analog of the conventional (attractive) Casimir force can be found in our formalism as a nonadiabatic correction of order $\order{\hbar\omega_p/\Delta}$, and thus small compared to $E_{\rm QG}$ in the low frequency regime (see SM for more details).

The zero-point energy $E_{\rm QG}$ has another measurable consequence.
Since $E_{\rm QG}$ depends on $C_p$, it induces an attractive force acting on the capacitor of the circuit, which we also call the quantum geometric force. 
For a capacitor made of two parallel metal plates of area $S$ separated by distance $d$, the capacitance is given by $C_p = \epsilon S/d$ where $\epsilon$ is the dielectric constant of the medium between the plates.
Then, the force acting on the capacitor in a circuit state with $n_p$ photons is given by $F\equiv\pdv*{E_{n_p}}{d}$ where $E_{n_p}$ is the total energy with $n_p$ photons, yielding
\begin{align}
	F &= F_{n_p} + \frac{e^2}{2\epsilon S} \vec{g} G \vec{g} = F_{n_p} + F_{\rm QG}. \label{eq:quantum_geometric_force1}
\end{align}
Here, $\expval{\dots}$ is the expectation value in the ground state, and a positive $F$ means the attractive force between the capacitor plates. $F_{n_p} = \hbar\omega_p (n_p + 1/2)/(2d)$ is the force without the electronic system due to the $C_p$ dependence of the photon frequency $\omega_p$.

Eq.~\eqref{eq:quantum_geometric_force1} is the second main result of this work. The second term $F_{\rm QG}$ in Eq.~\eqref{eq:quantum_geometric_force1} represents the force arising from the electronic system coupled to fluctuating electromagnetic fields. It is proportional to the many-body quantum metric of the electronic system, and does not depend on the state of photons in the circuit. 
Even when the system is at finite temperature $T$, the temperature only affects $F_p$ and does not affect $F_{\rm QG}$, as long as the temperature $T$ is low compared to the electronic system's energy gap. The quantum geometric force is controlled by the ground state property of the electronic system, the many-body quantum metric $G$.

The dependence of $F_{\rm QG}$ on the parameters of the capacitor should be highlighted. It does not depend on the capacitor plate distance $d$, and is inversely proportional to the plate area $S$ as $F_{\rm QG}\propto d^0S^{-1}$. This is in sharp contrast to the conventional Casimir force, which typically scales as $\propto d^{-4}S^{+1}$~\cite{casimir1948attraction}.  These dependencies on the capacitor parameters, along with the dependence on the electronic system, make the quantum geometric force distinct from the conventional Casimir force.

In general, other effects can contribute to the force on the capacitor $F$~\cite{shahmoon2018,barashCasimirLifshitzInteractionBodies2026}. However, at large $d$, we expect the quantum geometric force to dominate, because it arises from the mode that couples directly to the capacitor charge $\hat{q}$. Other electromagnetic modes not considered here, such as vacuum fluctuations, may also contribute to $F$, but they couple only to the dipole and higher multipole moments of each capacitor plate. Therefore, their contributions become subleading at large $d$. The Casimir force discussed above is one such example.

Importantly, the energy shift $E_{\rm QG}$ and the resulting forces can occur even in a macroscopic system that involves $\sim10^{23}$ electrons. They are extensive in the system volume $\Vol$, because the quantum metric $G_{ij}$ for a macroscopic gapped system is extensive. For example, consider an atomic insulator where electrons are tightly bound with an orbital spread $\Delta R$ much smaller than the lattice constant $a$ ($\Delta R \ll a$) and hopping between orbitals is negligible. Since such a system is essentially a collection of independent atoms, the quantum metric $G$ reduces to the sum of quantum metrics of the individual atoms and is extensive. 

More generally, for band insulators in $d$-dimensions, the many-body quantum metric $G$ is independent of $\vec{A}$ and can be expressed in terms of  
the quantum metric of Bloch wavefunctions at each $\vec{k}$ point~\cite{marzari1997,souza2000,onishi2025b}: 
\begin{align}
    G_{ij} &= \Vol\Re\int\frac{\dd^d{k}}{(2\pi)^d} \mel{\partial_{k_i} u(\vec{k})}{(1-P(\vec{k}))}{\partial_{k_j} u(\vec{k})} \label{eq:K}
\end{align}
where $P(\vec{k}) = \ketbra{u(\vec{k})}{u(\vec{k})}$, $\ket{u(\vec{k})}=|u_1(\vec{k})\dots u_r(\vec{k})|$ is the Slater determinant state of all the $r$ occupied Bloch states at $\vec{k}$. The corresponding intensive quantity in this case, 
\begin{align}
    K_{ij} &= \frac{2\pi}{\Vol}G_{ij}, \label{K-G}
\end{align}
is a ground state property that characterizes long-wavelength density fluctuation, recently termed as quantum weight~\cite{onishi2025b}.

For real solids where electrons interact with long-range Coulomb forces, the effect of dielectric screening needs to be considered in understanding the change of the many-body ground state with the vector potential \cite{onishi2025b}. 
On the other hand,  for two-dimensional materials with an energy gap, because dielectric screening is absent at long wavelength ($q\rightarrow 0$), the relation \eqref{K-G} between many-body quantum metric $G$ and the quantum weight $K$ continues to hold, where $K$ is defined in terms of the static structure factor in the $q\rightarrow 0$ limit~\cite{onishi2025b}: 
\begin{align}
    K_{ij} &= \frac{\pi}{\Vol} \eval{\pdv{\expval{n_{\vec{q}}n_{-\vec{q}}}}{q_iq_j}}_{\vec{q} \rightarrow 0}, \label{K-S}
\end{align}
where $n_{\vec{q}}=\int\dd{\vec{r}}e^{-i\vec{q}\cdot\vec{r}}n(\vec{r})$ is the number density operator with wavevector $\vec{q}$. 

A nontrivial example is a (fractional) Chern insulator with a quantized Hall conductivity~\cite{Thouless1982,niu1985}. %
The quantum weight $K$ in (fractional) Chern insulators has a topological bound set by the Chern number $\mathcal{C}$: $\Tr K\ge \abs{\mathcal{C}}$~\cite{onishi2024c}. Another interesting case is one- and two-dimensional systems near gap-closing transitions, at which the quantum weight generally diverges ~\cite{onishi2024e}. While our theory breaks down when the gap is smaller than the circuit frequency $\sim\SI{10}{\micro\electronvolt}$, we expect an enhancement in the zero-point energy $E_{\rm QG}$ as we approach gap-closing transitions.

By measuring the quantum geometric force $F_{\rm QG}$, we can directly probe the quantum geometry of solids. Previously, $k$-space single-particle quantum metric was measured in solids with two-band effective Hamiltonians~\cite{kang2024,kim2025}. For more general systems, methods relying on sum rules requiring broad-frequency measurements~\cite{souza2000,onishi2025b,ghosh2024}, or equivalent time domain protocols~\cite{verma2025a} were proposed, and sum-rule–based measurements of quantum weight have been reported recently~\cite{balut2025a,balut2026}. In contrast, the quantum geometric force  provides a direct and static probe of the quantum weight in general gapped systems, including interacting systems.

\begin{figure}[htbp]
	\centering
	\includegraphics[width=0.9\columnwidth]{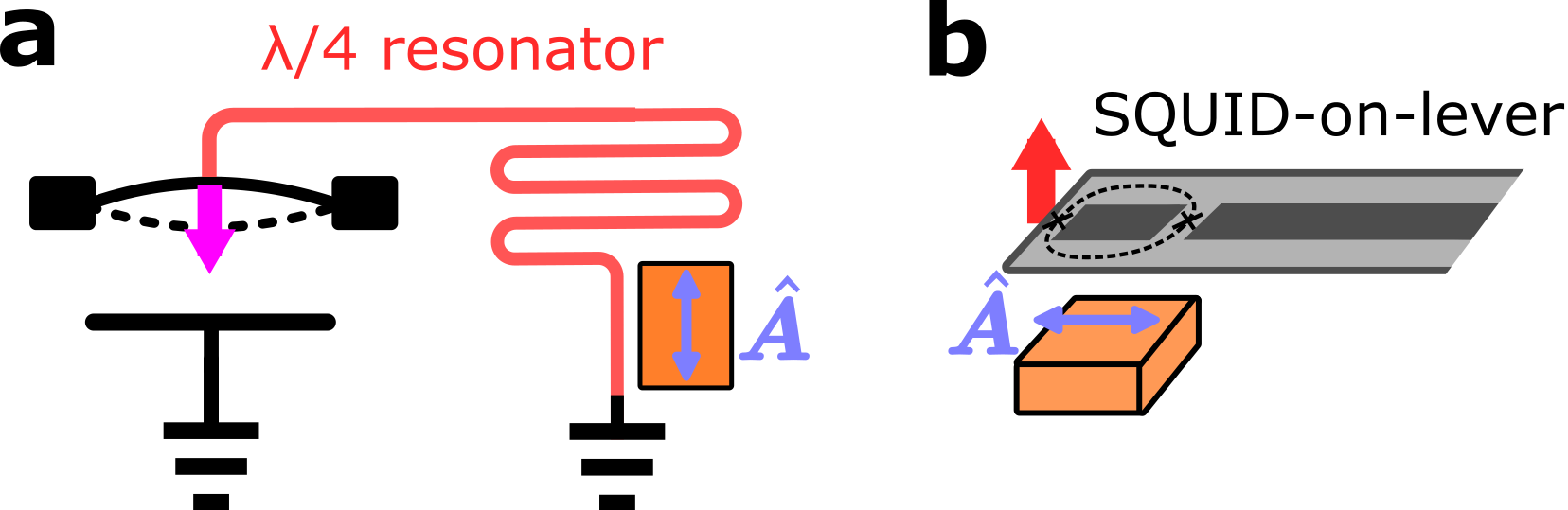}
	\caption{A possible experimental realization. (a) Measurement of the attractive quantum geometric force. The capacitor in the circuit is made of a mechanical resonator, so that it can move and detect the quantum geometric force. (b) Measurement of the repulsive quantum geometric force. SQUID is fabricated on a cantilever (SQUID-on-lever). SQUID acts as a LC circuit approximately, and the quantum fluctuation in SQUID induces the repulsive quantum geometric force when it is close contanct to the material. }
	\label{fig:cylinder}
\end{figure}

\textit{Experimental setup. ---} Possible experimental setups are shown in Fig.~\ref{fig:cylinder}.
To measure the attractive quantum geometric force, we can use a capacitor made of a microelectromechanical resonator~\cite{xu2022} as shown in Fig.~\ref{fig:cylinder}(a). The capacitor plates then can move and thus detect the quantum geometric force acting on it. A superconducting circuit coupled to a micromechanical resonator was previously realized in experiments~\cite{teufel2011,reed2017}.
To realize large inductive coupling, we can place the material near the grounded end of the $\lambda/4$ resonator forming the LC circuit, where the fluctuating current is large and relatively uniform. We can also deposit the superconducting wire on the material to maximize the coupling, as done in previous experiments~\cite{tanaka2025}.

We estimate the size of $\vec{g}$ in this case with a typical mutual inductance $M$ as $\vec{g}\sim M/(l L_p)\sim\SI{0.1}{\per\milli\metre}$ for $l\sim \SI{1}{\milli\metre}$, $M\sim\SI{1}{\nano\henry}$ and $L_p\sim\SI{10}{\nano\henry}$. For the system volume $\Vol=\SI{100}{\milli\meter^3}$, the quantum weight $K\sim\SI{1}{\per\angstrom}$, and the capacitance $C_p=\SI{1}{\femto\farad}$, the zero-point energy is $E_{\rm QG}\sim \SI{1}{\electronvolt}$. For the capacitor made of two parallel plates with area $S\simeq\SI{100}{\micro\meter^2}$ separated by a vacuum gap of distance $d\sim\SI{900}{nm}$, we estimate $F_{\rm QG} \sim \SI{0.2}{\pico\newton}$. Both $E_{\rm QG}$ and $F_{\rm QG}$ are significantly large compared to the typical energy scale of a superconducting circuit $\hbar\omega_p \sim \SI{200}{\micro\electronvolt}$ and the photon-induced force $F_{n_p}\sim \SI{0.06}{\femto\newton}$ for $n_p\sim 1$. The quantum geometric force $F_{\rm QG}$ is within the measurable range of current experimental techniques of microelectromechanical systems, where force sensitivity of $\SI{390}{zN~\hertz^{-1/2}}$ has been achieved~\cite{weber2016}. 

We note that the quantum geometric force can be comparable to the conventional Casimir force~\cite{klimchitskaya2009}. The standard Casimir force between two parallel metal plates is given by $F_{c}=K_cS/d^4$ with $K_c=\pi h c/480=\SI{1.3e-27}{Nm^2}$ and the plate distance $d$ and the area $S$. Since $F_{\rm QG}\propto S^{-1}d^0$, $F_{\rm QG}$ can become dominant over $F_{\rm c}$ for small $S$ and large $d$. In the above estimation, the Casimir force is estimated to be $F_c\sim \SI{0.2}{pN}$, comparable to the quantum geometric force.

For measurement of the repulsive quantum geometric force, we can use an SQUID instead of the LC circuit as shown in Fig.~\ref{fig:cylinder}(b). We can consider the Josephson junction in the SQUID as a (nonlinear) inductor with an effective capacitance, %
thus acting as an LC circuit.
We note that the zero-point energy~\eqref{eq:E_QG} does not change for nonlinear inductor such as Josephson junction. 
By measuring the force acting on the SQUID, we can measure the repulsive quantum geometric force. Such a force measurement may be realized by the SQUID-on-Lever technique~\cite{wyss2022,weber2025}, where a SQUID is fabricated on the cantilever of the atomic force microscopy. 

In this case, we estimate the coupling constant $\vec{g}$ using the formula for the magnetic moment in magnetostatics: $\vec{A}\sim \mu_0 I A/(4\pi R^2)$, where $\mu_0$ is the vacuum permeability, $I$ is the current in the probe, $A$ is the area enclosed by the current path, and $R$ is the distance between the probe and the target system. This yields $g \sim \mu_0 A/(4\pi R^2 L_p)\sim\SI{10}{\per\milli\metre}$ for $A\sim \SI{1}{\micro\meter\squared}$, $R \sim \SI{100}{\nano\meter}$ and $L_p\sim\SI{10}{\nano\henry}$. For the capacitance $C_p=\SI{10}{\femto\farad}$ and a system with volume $\Vol\sim\SI{1}{\micro\meter^3}$, $K\sim\SI{1}{\per\angstrom}$, the zero-point energy is $E_{\rm QG}\sim \SI{1}{\micro\electronvolt}$ and the resulting repulsive quantum geometric force is $\nabla E_{\rm QG} \sim E_{\rm QG}/R \sim \SI{1}{\femto\newton}$. 

We note that the measured force between the circuit or SQUID and the material will include other contributions, such as conventional Casimir forces, making the direct observation of the quantum geometric contribution challenging. In this regard, the quantum geometric force on the capacitor plates can be comparable to the other contributions in the large-$d$ and small-$S$ limit, hence more promising for experimental observations.

In both cases, the predicted forces depend on the material. When we tune the material's quantum weight by, e.g., gating, the forces should change accordingly. The expected change can be significant, especially near the gap-closing phase transitions in reduced dimensions as mentioned above. The dependence on the material property is one of the key signatures of the quantum geometric force, distinguishing it from other sources. 

Lastly, we note that our theory also applies to two coupled superconducting circuits as discussed in SM.

\textit{Discussion. ---} The zero-point energy $E_{\rm QG}$ and the resulting forces are essentially ground state phenomena. They can occur even when the circuit does not excite any real photons. %
Only through virtual photons, the electrons in solids induce the zero-point energy $E_{\rm QG}$ and results in the repulsive and attractive quantum geometric forces. 
Our discussion also shows that the many-body quantum metric naturally emerges when both the electronic systems and the electromagnetic fields are quantum mechanical. 
This is in sharp contrast to the previously discussed effect associated with the Berry curvature~\cite{sundaram1999,Xiao2010}, or more generally the quantum geometry~\cite{gao2014,gao2023,wang2023b,souza2000,Ahn2020,ahn2022,onishi2024e,onishi2024c,ghosh2024,verma2025,verma2025a,komissarov2024,neupert2013}, 
which treated electromagnetic fields classically. 

Similar phenomena may occur in other setups beyond the superconducting LC circuit. 
For example, an optical cavity also has a fluctuating electromagnetic field which can be coupled to quantum materials~\cite{haroche2006}. Even when the cavity is in the vacuum state, the quantum fluctuation of the electromagnetic field inside the cavity would induce the zero-point energy $E_{\rm QG}$ similar to the superconducting circuit case, although it can involve multiple modes with different frequencies in the cavity.

The effects proposed in this work highlight a new form of interplay between quantum materials and superconducting circuits, and suggest these circuits as a powerful approach to probe and exploit fundamental properties of quantum materials.

\acknowledgements
We thank Miuko Tanaka, Joel \^{I}-j. Wang, Leonid Glazman and Daniil Antonenko for useful discussions.
YO was supported in part by Grant No. NSF PHY-1748958 to the Kavli Institute for Theoretical Physics (KITP), the Heising-Simons Foundation, and the Simons Foundation (216179, LB). LF was supported by the Air Force Office of Scientific Research under award number FA2386-241-4043.

\bibliography{references}

\appendix
\clearpage
\newpage

\begin{widetext}
\section*{Supplemental Materials}
\input{SM_content}

\end{widetext}

\end{document}

%% file: preamble.tex
\usepackage[utf8]{inputenc}

\usepackage{amsmath,amssymb,ascmac,fancybox}
\usepackage{color}
\usepackage{graphicx}
\usepackage{url}
\usepackage{siunitx}
\usepackage{bm}
\usepackage{physics}
\usepackage{mathtools}
\usepackage{hyperref}
\usepackage[T1]{fontenc}
\usepackage{ulem}
\usepackage{docmute} %

\graphicspath{figures/}

\newcommand{\ani}[1]{\hat{#1}}

\renewcommand{\vec}{\vb*}

\newcommand{\Vol}{\Omega}

%% file: SM_content.tex
\title{Supplemental Materials for ``Zero-point energy of solids from vacuum fluctuation and quantum geometric force''}
\author{Yugo Onishi}
\affiliation{Department of Physics, Massachusetts Institute of Technology, Cambridge, MA 02139, USA}
\author{Liang Fu}
\affiliation{Department of Physics, Massachusetts Institute of Technology, Cambridge, MA 02139, USA}
\date{\today}

\maketitle

\section{Equivalence with the many-body quantum geometric tensor defined through the twisted boundary condition}

In the main text, we define the many-body quantum metric as the change of the ground state when a uniform vector potential changes. We show in the main text that, in open-boundary systems, the quantum metric defined this way reduces to the polarization fluctuation. 

Here we show that, for the periodic boundary condition, the quantum metric defined in the main text reduces to the quantum metric defined with the twisted boundary condition, defined in literature~\cite{niu1985,souza2000}.

The many-body quantum metric of the ground state is defined with the twisted boundary condition~\cite{niu1985}, which imposes the many-body wavefunction to satisfy the following boundary condition:
\begin{align}
	&\Psi_{\vec{\kappa}}(\dots, \vec{r}_{j-1}, \vec{r}_j + L\hat{x}, \vec{r}_{j+1}, \dots) \nonumber \\
	&= e^{i\kappa_x L}\Psi_{\vec{\kappa}}(\dots, \vec{r}_{j-1}, \vec{r}_j, \vec{r}_{j+1}, \dots),
\end{align}
where $L$ is the system size and $\kappa_x$ specifies the twist angle along $x$ direction, and similarly for $y$ and $z$ directions. Here, $\vec{r}_j$ represents the coordinate of the $j$-th particle. When $\vec{\kappa}=0$, the twisted boundary condition reduces to the periodic boundary condition. To define the many-body quantum geometry, we need an analog of cell-periodic Bloch wavefunction for the many-body wavefunction, which we denote by $\Phi_{\vec{\kappa}}$ and is defined as
\begin{align}
	\Phi_{\vec{\kappa}}(\vec{r}_1, \vec{r}_2, \dots, \vec{r}_N) = e^{-i\vec{\kappa}\cdot\sum_j \vec{r}_j}\Psi_{\vec{\kappa}}(\vec{r}_1, \vec{r}_2, \dots, \vec{r}_N).
\end{align}
The wavefunction $\Phi_{\vec{\kappa}}$ satisfies the periodic boundary condition: $\Phi_{\vec{\kappa}}(\dots, \vec{r}_j + L\hat{x}, \dots) = \Phi_{\vec{\kappa}}(\dots, \vec{r}_j, \dots)$ and similar for the other directions. $\Phi_{\vec{\kappa}}$ is the eigenstate of the $\vec{\kappa}$-dependent Hamiltonian 
	$H(\vec{\kappa}) = e^{-i\vec{\kappa}\cdot\sum_j \vec{r}_j} H e^{i\vec{\kappa}\cdot\sum_j \vec{r}_j}$.

The many-body quantum geometric tensor is then defined as
\begin{align}
	Q_{ij} = \frac{2\pi}{V}\mel{\pdv{\Phi_0}{\kappa_i}}{(1-P_0)}{\pdv{\Phi_0}{\kappa_j}} = K_{ij} - \frac{i}{2}C_{ij}. \label{eq:Q_def}
\end{align}
Here, $P_0 = \ketbra{\Phi_0}{\Phi_0}$ is the projection operator onto the ground state and $V$ is the volume of the system. The real part of $Q_{ij}$ defines the many-body quantum metric $K_{ij} = \Re Q_{ij}$, while the imaginary part defines the many-body Berry curvature $F_{ij} = 2\Im C_{ij}$ and is related to the Chern number $\mathcal{C}$ in two-dimensional systems as $C_{xy} = \mathcal{C}$. The quantum metric in this particular context was recently termed as quantum weight~\cite{onishi2025b}, which we shall also use in this paper.

The quantum weight $K_{ij}$ quantifies the change of the ground state under the change of the twisted boundary condition. Physically, it can be interpreted with an external magnetic flux. When the system forms a ring with circumference $L$ implementing the periodic boundary condition and the magnetic flux $\Phi$ is threaded through the ring, the magnetic flux produces a uniform vector potential $A=\Phi/L$ along the ring. This geometry with uniform vector potential $\vec{A}$ is equivalent to the twisted boundary condition with $\vec{\kappa}$ related by
\begin{align}
	\vec{A} = \frac{\hbar}{e}\vec{\kappa}, \label{eq:A_kappa}
\end{align}
with $e(<0)$ the charge of an electron. Therefore, the quantum geometric tensor defined with the twisted boundary condition is equivalent to that defined through the vector potential as in the main text, up to the factor in Eq.~\eqref{eq:A_kappa}.

\section{Effect of capacitive coupling}
In this section, we consider the effect of the capacitive coupling, i.e., the coupling through the scalar potential between the electronic system and the superconducting circuit. 

For purely capacitive coupling case with $\vec{g}_I=0$, we can simply treat the capacitive coupling as a perturbation to derive a negative energy shift and an attractive force between the electronic system and the circuit. However, this treatment is not clear how it is related to the adiabatic treatment presented in the main text when both the capacitive and inductive couplings are present. 

Here, we present a more general treatment that can handle both couplings on equal footing. We show that, even in the presence of the capacitive coupling, the positive energy shift and the repulsive force originating from the quantum geometry dominates in the adiabatic limit $\hbar\omega_p \ll \Delta$ with the gap of the electronic system $\Delta$. Then, we also show that a negative energy shift and an attractive force appears as the nonadiabatic correction, which is suppressed by the factor $\hbar\omega_p/\Delta$ compared to the adiabatic effect. The resulting negative energy shift and the attractive force can be interpreted as the analog of the van der Waals force in our system.

\subsection{Effective Hamiltonian}
We consider the following total Hamiltonian including both the vector and scalar potential couplings:
\begin{align}
H_{\rm tot} &= H[\vec{A}(\vec{r})] + \int\dd^dr\,\rho(\vec{r})\phi(\vec{r}) + H_p, \label{eq:Htot_spatial}
\end{align}
where $H[\vec{A}(\vec{r})]$ is the Hamiltonian of the electronic system in the presence of the vector potential $\vec{A}(\vec{r})$, $\rho(\vec{r})$ is the charge density operator of the electronic system, and $\phi(\vec{r})$ is the scalar potential generated by the circuit. $H_p$ is the Hamiltonian of the circuit, which we model as a single-mode LC resonator as in the main text:
\begin{align}
	H_p = \frac{\hat{\varphi}^2}{2L_p} + \frac{\hat{q}^2}{2C_p},
\end{align}
where $\hat{\varphi}, \hat{q}$ are the flux and charge operators of the resonator, satisfying $\comm{\hat{\varphi}}{\hat{q}} = i\hbar$. The vector and scalar potentials generated by the circuit are given by
\begin{subequations}
\begin{align}
	\hat{\phi}(\vec{r}) &= g_C(\vec{r}) \frac{\hat{q}}{C_p}, \label{ap:eq:phi_gC} \\
	\hat{\vec{A}}(\vec{r}) &= \vec{g}_I(\vec{r}) \hat{\varphi},
\end{align}
\end{subequations}
where $g_C(\vec{r}), \vec{g}_I(\vec{r})$ are the spatial profiles of the scalar and vector potentials generated by the circuit per unit flux/charge. 

Before proceeding, we note that the couplings through the scalar and vector potential are related to each other by a gauge transformation. To see this, we consider the following unitary transformation:
\begin{align}
	U_1 = \exp(-\frac{i\hat{\varphi}}{\hbar}\int\dd^dr g_C(\vec{r}) \rho(\vec{r})).
\end{align}
This unitary transformation induces the gauge transformation for the electrons in the target that translates the scalar potential into the vector potential. For example, the annihilation operator of an electron $\ani{\psi}$ transforms under $U_1$ as 
	$U_1^{\dagger}\ani{\psi}(\vec{r})U_1 = e^{-ie g_C(\vec{r})\hat{\varphi}/\hbar} \ani{\psi}(\vec{r})$,
where $e(<0)$ is the charge of an electron.
Therefore, $U_1$ induces the gauge transformation of the vector potential as $\vec{A}\to \vec{A}+\hat{\varphi}\nabla g_C(\vec{r})$. On the other hand, $U_1$ shifts the charge operator $q$ of the probe as 
\begin{align}
	&U_1^{\dagger} q U_1 = q - Q_0, \\
	&Q_0 \equiv \int\dd^dr g_C(\vec{r}) \rho(\vec{r})
\end{align}
where $Q_0$ represents the charge in the probe induced through the interaction with the target.

With the unitary transformation $U_1$, the Hamiltonian~\eqref{eq:Htot_spatial} is transformed into 
\begin{align}
	&H'_{\rm tot} \equiv U_1^{\dagger}H_{\rm tot}U_1 = H'[\vec{A}(\vec{r})=\vec{g}(\vec{r})\hat{\varphi}] + H_p, \label{ap:eq:tot_H1}\\
	&H'[\vec{A}(\vec{r})] = H[\vec{A}(\vec{r})] - \frac{Q_0^2}{2C_p}, \\
	&\vec{g}(\vec{r}) = \vec{g}_I(\vec{r}) + \nabla g_C(\vec{r}).
\end{align}
Here, $H'$ is the Hamiltonian in the presence of the probe. The charge of the probe partially screens that of the target by $Q_0$, reducing the effective interaction within the target by $Q_0^2/(2C_p)$. In the form of Eq.~\eqref{ap:eq:tot_H1}, all the interaction effects between the target and probe are through the vector potential $\vec{A}(\vec{r})=\vec{g}(\vec{r})\varphi$ and the screening effect represented by $Q_0^2/(2C_p)$. 

\subsection{Quantum geometric force including the effect of charge coupling}
Here we consider the quantum geometric force including the effect of charge coupling. As discussed in the previous section, the screening effect modifies the Hamiltonian of the electronic system as
\begin{align}
	H' = H - \frac{Q_0^2}{2C_p}
\end{align}
where the last term represents the screening of the interaction due to the probe. Here, we show that the force on the capacitor of the circuit due to the screening is mostly canceled when the scalar potential generated by the charge $\hat{q}$ on the capacitor is given by Eq.~\eqref{ap:eq:phi_gC} and $g_C$ does not depend on $C_p$. This is true when the electronic system is placed between the capacitor plates so that the scalar potential on the electronic system is proportional to the electric field generated by the capacitor, $\vec{E}\propto \hat{q}/C_p$.

When the capacitor is made of two parallel metal plates with area $S$ and gap size $d$, the capacitance is given by $C_p = \epsilon S/d$ and the force acting on the capacitor is given by 
\begin{align}
	F &= \pdv{E_{n_p}}{d} = F_{n_p} + \frac{V}{2\epsilon S} \qty(-\frac{\expval{Q_0^2}}{V} + \frac{e^2}{2\pi} \vec{g} K' \vec{g}) \label{ap:eq:quantum_geometric_force}
\end{align}
Here, $\expval{\dots}$ is the expectation value in the ground state, and $F_{n_p} = \hbar\omega_p (n_p + 1/2)/(2d)$ is the force without the interaction with the target system, which arises from the dependence of the resonant frequency $\omega_p$ on $d$.

Now we show that $\expval{Q_0^2}$ is related to $K'$. We first rewrite $Q_0$ as 
\begin{align}
	Q_0 &= \int\dd^dr g_C(\vec{r}) \rho(\vec{r}) = \frac{1}{V}\sum_{\vec{q}} g_C(-\vec{q}) \rho(\vec{q})
\end{align}
where $\rho(\vec{q})=\int\dd^dr e^{-i\vec{q}\vdot\vec{r}} \rho(\vec{r}), g_C(\vec{q}) = \int\dd^dr e^{-i\vec{q}\vdot\vec{r}} g_C(\vec{r})$ are the Fourier transform of the charge density operator $\rho(\vec{r})$ and $g_C(\vec{r})$, and $V$ is the volume of the system. Using this, we find that $\expval{Q_0^2}$ is given by
\begin{align}
	\expval{Q_0^2} &= \frac{1}{V^2} \sum_{\vec{q},\vec{q}'} g_C(-\vec{q}) g_C(-\vec{q}') \expval{\rho(\vec{q})\rho(\vec{q}')} \nonumber \\
	&= \expval{Q_0}^2 + \frac{1}{V} \sum_{\vec{q}} \abs{g_C(\vec{q})}^2 S(\vec{q}) \label{ap:eq:Q0_structurefactor}
\end{align}
where $S(\vec{q}) = (1/V)(\expval{\rho(\vec{q})\rho(-\vec{q})}-\expval{\rho(\vec{q})}\expval{\rho(-\vec{q})})$ is the static structure factor. Here we assumed that $g_C(\vec{r})$ is slowly varing in space so that $g_C(\vec{q})$ has significant weight only near $\vec{q}=0$, and the system is translationally invariant so that $\expval{\rho(\vec{q})\rho(\vec{q}')}$ is finite only when $\vec{q}+\vec{q}'$ coincides with a reciprocal lattice vector. 

Since the static structure factor is related to the quantum weight as $S(\vec{q}) = e^2 \vec{q} K' \vec{q}/(2\pi)$ for small $\vec{q}$~\cite{onishi2025b}, we find that 
\begin{align}
	\expval{Q_0^2} &= \expval{Q_0}^2 + \frac{e^2}{2\pi V} \sum_{\vec{q}} \abs{g_C(\vec{q})}^2 q_i K'_{ij} q_j \nonumber \\
	&= \expval{Q_0}^2 + \frac{e^2}{2\pi}\int\dd^dr (\nabla g_C) K' (\nabla g_C) \\
	&= \expval{Q_0}^2 + \frac{e^2 V}{2\pi} (\nabla g_C) K' (\nabla g_C)
\end{align}
where in the last line we assumed that $g_C(\vec{r})$ varies slowly in space so that we can approximate $\nabla g_C$ as constant. Plugging this into the expression for $F$~\eqref{ap:eq:quantum_geometric_force}, we obtain
\begin{align}
	F &= F_{n_p} - \frac{\expval{Q_0}^2}{2\epsilon S} + \frac{e^2 V}{4\pi \epsilon S} \qty(-(\nabla g_C) K' (\nabla g_C) + \vec{g} K' \vec{g}) \\
	&= F_{n_p} - \frac{\expval{Q_0}^2}{2\epsilon S} + \frac{e^2 V}{4\pi\epsilon S} \qty(\vec{g}_I K' \vec{g}_I + 2(\nabla{g}_C) K' \vec{g}_I)
\end{align}
Lastly, we note that $Q_0$ includes all the charges in the target, including the positive background charge. Therefore, if $g_C(\vec{r})$ is slowly varying in space compared to the lattice constant of the target, we have $\expval{Q_0}\simeq0$. Thus, the force acting on the capacitor is given by 
\begin{align}
	F &= F_{n_p} + \frac{e^2 V}{4\pi \epsilon S} \qty(\vec{g}_I K' \vec{g}_I + 2(\nabla{g}_C)K' \vec{g}_I)
\end{align}

\subsection{Nonadiabatic correction}
Here we consider nonadiabatic corrections to the discussion in the main text. We start with the following Hamiltonian:
\begin{align}
	H_{\rm tot} &= H(\vec{A}=\vec{g}\hat{\varphi}) + H_p, \\
	H_p &= \frac{\hat{\varphi}^2}{2L_p} + \frac{\hat{q}^2}{2C_p},
\end{align}
We assume $\vec{g}$ is uniform and does not depend on position $\vec{r}$. We can include the effect of the scalar potential coupling by redefining $\vec{g}$ and $H(\vec{A})$ as discussed in the previous section. 

We denote the unitary transformation that diagonalizes $H(\vec{A})$ by $U$. $U$ satisfies $U^{\dagger}(\vec{A}) H(\vec{A}) U(\vec{A}) = \sum_n E_n(\vec{A}) \ketbra{n}{n}$ with $\ket{n}$ the eigenstate of $H(\vec{A})$. By applying $U(\vec{g}\hat{\varphi})$ to $H_{\rm tot}$, we obtain
\begin{align}
	\tilde{H} \equiv U^{\dagger} H_{\rm tot} U &= \sum_n E_n(\vec{g}\hat{\varphi}) \ketbra{n}{n} + \frac{\hat{\varphi}^2}{2L_p} + \frac{(\hat{q} - \hbar\vec{g}\vdot\vec{a})^2}{2C_p}
\end{align}
where $\vec{a}_{mn}(\vec{A}) = i\mel{m}{\nabla_{\vec{A}}}{n}$ is the non-Abelian Berry connection. We write $\tilde{H}$ as $\tilde{H} = H_{\rm ad} + V$ where $H_{\rm ad}$ and $V$ is given by 
\begin{align}
	&H_{\rm ad} = \sum_n (E_n(\vec{g}\hat{\varphi}) + E_{{\rm QG}, n}) \ketbra{n}{n} + \frac{\hat{\varphi}^2}{2L_p} + \frac{(\hat{q}-Q)^2}{2C_p}, \\
	&V = -\frac{\hbar (\vec{g}\vdot\vec{a})}{C_p} \hat{q} + \order{g^2}.
\end{align}
Here, $Q = \hbar\vec{g}\vdot\vec{a}_{\rm diag}$ is proportional to the diagonal part of the Berry connection denoted by $\vec{a}_{\rm diag}$, representing the induced charge in the circuit by the electronic system.  
$H_{\rm ad}$ is the adiabatic Hamiltonian and diagonal in the eigenbasis of $H(\vec{A})$, and it includes the quantum geometric potential for $n$-th eigenstate:
\begin{align}
	E_{{\rm QG}, n} &= \frac{e^2}{2C_p} \vec{g} G_n \vec{g}
\end{align}
with $(G_n)$ the many-body quantum metric for the $n$-th eigenstate. $V$ is the off-diagonal in the eigenbasis of $H(\vec{A})$, representing the nonadiabatic correction that induces transitions between different eigenstates of $H(\vec{A})$. In $V$, we have neglected the commutator between $\hat{q}$ and $\vec{a}$, which gives higher-order corrections in $\vec{g}$.
 
We treat $V\propto \vec{g}$ as a perturbation to $H_{\rm ad}$. Up to the second order in $\vec{g}$, we can derive the effective Hamiltonian for the ground state with the Schrieffer-Wolff transformation as
\begin{align}
	H_{\rm eff} &= E_0(\vec{g}\hat{\varphi}) + E_{{\rm QG}, 0} + \frac{\hat{\varphi}^2}{2L_p} + \frac{(\hat{q}-q_0)^2}{2C_p} + \Delta H_{\rm NA} + \order{g^3}, \\
	\Delta H_{\rm NA} &= - \sum_{n\neq0} \frac{\hbar^2}{C_p^2} \frac{(\vec{g}\vdot\vec{a}_{0n})(\vec{g}\vdot\vec{a}_{n0})}{E_n - E_0} \hat{q}^2
\end{align}
where $q_0 = \hbar\vec{g}\vdot\vec{a}_{00}$ and $E_n$ is the $n$-th eigenvalue of $H$. $\Delta H_{\rm NA}$ is the nonadiabatic correction to the effective Hamiltonian, which arises from the virtual excitation to the excited states due to the nonadiabatic coupling $V$. Noting that the eigenstate under uniform $\vec{A}$ is related to that at $\vec{A}=0$ as $\ket{n(\vec{A})} = e^{-(ie/\hbar)\vec{A}\vdot\vec{R}}\ket{n(0)}$ with $\vec{R}=\sum_i \vec{r}_i$ the total position operator, we have $\vec{a}_{mn} = -(e/\hbar)\mel{m}{\vec{R}}{n}$. Using this, we can rewrite $\Delta H_{\rm NA}$ as
\begin{align}
	\Delta H_{\rm NA} &= - \frac{e^2 \hat{q}^2}{C_p^2} g_i g_j \sum_{n\neq0}  \frac{\Re[\mel{0}{R_i}{n}\mel{n}{R_j}{0}]}{E_n - E_0} 
\end{align}

We can show that $\Delta H_{\rm NA}$ can be rewritten with the static polarizability $\chi_{ij}$ of the electronic system. The polarizability $\chi_{ij}$ is defined as 
\begin{align}
	\vec{p} &= \chi \vec{E}, 
\end{align}
where $\vec{E}$ is the applied electric field and $\vec{p}$ is the induced polarization density. Note that $\chi$ is an intensive quantity. Using the Kubo formula, we can express $\chi_{ij}$ as
\begin{align}
	\chi_{ij} &= \frac{2 e^2}{\Vol}\sum_{n\neq0} \frac{\Re[\mel{0}{R_i}{n}\mel{n}{R_j}{0}]}{E_n - E_0}
\end{align}
where $\Vol$ is the system volume. Comparing this with $\Delta H_{\rm NA}$, we find 
\begin{align}
	\Delta H_{\rm NA} &= - \frac{\Omega}{2}\frac{\hat{q}^2 \vec{g}\chi\vec{g}}{C_p^2}.
\end{align}

$\Delta E_{\rm NA}$ can be interpreted as the energy gain due to the dielectric response of the material to the fluctuating electric field generated by the circuit. The electric field generated by the circuit is given by $\hat{\vec{E}} = -\pdv*{\hat{A}}{t} - \nabla\hat{\phi}$, which yields
\begin{align}
	\hat{\vec{E}} &= -\vec{g}_I \dot{\hat{\varphi}} - \nabla g_C \frac{\hat{q}}{C_p} = -\vec{g}\frac{\hat{q}}{C_p}
\end{align}
With this, we can rewrite the nonadiabatic correction as
\begin{align}
\Delta H_{\rm NA} = - \frac{\Vol}{2}\hat{\vec{E}}\chi\hat{\vec{E}}
\end{align}
This is nothing but the energy gain due to the dielectric response of the material to the fluctuating electric field generated by the circuit.

Plugging this into $H_{\rm eff}$, we obtain
\begin{align}
	H_{\rm eff} &= E_0(\vec{g}\hat{\varphi}) + E_{{\rm QG}, 0} + \frac{\hat{\varphi}^2}{2L_p} + \frac{(\hat{q}-q_0)^2}{2C_{p,{\rm eff}}} + \order{g^3}, \\
	C_{p,{\rm eff}} &= C_p + \vec{g}\Vol\chi \vec{g}.
\end{align}
Here, the nonadiabatic correction renormalizes the capacitance of the circuit from $C_p$ to $C_{p,{\rm eff}}$. The renomarlization of the capacitance can be understood as the increased inertia of the flux in the circuit due to the electronic system. 

The eigenenergy of $H_{\rm eff}$ is given by
\begin{align}
	E_{n_p} &= E_0(\vec{g}\hat{\varphi}) + E_{{\rm QG}, 0} + \hbar\omega_{p,{\rm eff}}(n_p + 1/2) = E_0 +  E_{{\rm QG}, 0} + \hbar\omega_{p}(n_p + 1/2) + \delta E_{\rm NA} + \order{g^3}, \\
	\delta E_{\rm NA} &= - \hbar\omega_p(n_p + 1/2) \frac{\vec{g}\Vol\chi \vec{g}}{2C_p},
\end{align}
where $\omega_{p,{\rm eff}} = 1/\sqrt{L_p C_{p,{\rm eff}}}$ and $\omega_p=1/\sqrt{L_p C_p}$. $\delta E_{\rm NA}$ is the nonadiabatic correction to the eigenenergy. Importantly, the nonadiabatic correction $\delta E_{\rm NA}$ is always negative, and hence contribute to an attractive force between the circuit and the material.

$\delta E_{\rm NA}$ depends on the photon number $n_p$ in the resonator, in contrast to the adiabatic quantum geometric energy that is independent of $n_p$. In addition, the nonadiabatic correction vanishes when $L_p\to \infty$ and $\omega_p \to 0$, i.e., in the DC limit. This is in sharp contrast to the adiabatic quantum geometric energy that remains finite even in the DC limit. $\delta E_{\rm NA}$ involves the static polarizability $\chi$ of the electronic system in analogous to the van der Waals force or the usual Casimir effect. 

Since $\delta E_{\rm NA}$ is nonadiabatic correction, it scales differently with the energy gap of the material $\Delta$. Noting that the static polarizability roughly scales with the ground state quantum metric $G_0$ as $\Omega\chi \sim e^2 G_0/\Delta$, we find that the nonadiabatic correction $\delta E_{\rm NA}\sim -\hbar\omega_p (e^2/(2C_p))$ 
\begin{align}
	\delta E_{\rm NA} &\sim - (n_p + 1/2) \frac{\hbar\omega_p}{\Delta} \frac{e^2}{2C_p} \vec{g}G_0\vec{g} \sim - (n_p + 1/2) \frac{\hbar\omega_p}{\Delta} E_{\rm QG, 0} 
\end{align}
Therefore, the nonadiabatic correction is much smaller compared to the quantum geometric energy $E_{\rm QG, 0}$ when the energy gap $\Delta$ is much larger than the cavity frequency $\hbar\omega_p$. 

Even if the nonadiabatic correction is not negligible, we can, in principle, distinguish its effect from the quantum geometric force on the capacitor plate discussed in the main text. Since the nonadiabatic correction renormalizes the capacitance, its effect on the force on the capacitor plate is through the change of the circuit resonance frequency. By measuring the resonance frequency, we can extract the nonadiabatic effect from the experiment to distinguish the quantum geometric effect. In addition, for the force on the capacitor plates, the dependence on the plate distance $d$ is also different for the quantum geometric force and the nonadiabatic effect: the former is independent of $d$, while the latter depends on $d$ as $d^{1/2}$. These differences also help us distinguish the quantum geometric force from the nonadiabatic effect.

\section{Remark on systems with long-range Coulomb interaction}
In the presence of long-range Coulomb interaction, it was previously shown that the relation between the quantum weight and the structure factor can be modified due to the screening effect~\cite{resta2006,onishi2025b}. Therefore, a more careful discussion is needed to properly take into account the screening effect due to the Coulomb interaction. Here we show that our results will not change by properly taking into account the screening effect, and the quantum geometric force is still given by the same expression as in the main text.

To discuss this, we first generalize the quantum weight to finite wavevector $\vec{q}$. We consider the quantum system under an external vector potential with wavevector $\vec{q}$, $\vec{A}(\vec{r}) = 2\Re[\vec{A}_{\vec{q}} e^{i \vec{q} \cdot \vec{r}}]/V$. We denote the Hamiltonian under the vector potential by $H(\vec{A}_{\vec{q}})$, and the ground state by $\ket{\Psi_0(\vec{A}_{\vec{q}})}$. Then, we define the generalized many-body quantum metric as
\begin{align}
	K_{ij}(\vec{q}) &= 2\pi\qty(\frac{\hbar}{e})^2\Re\mel{\pdv{\Psi_0}{A_{\vec{q},i}}}{(1-P_0)}{\pdv{\Psi_0}{A_{\vec{q},j}}}. \label{eq:G_q_def}
\end{align}
where $P_0=\ketbra{\Psi_0}{\Psi_0}$ is the projection operator onto the ground state, and the derivatives are evaluated at $\vec{A}_{\vec{q}}=0$.

As discussed previously, the quantum weight behaves differently in the presence of long-range Coulomb interaction. This can be most clearly seen through its relation to optical responses at finite wavevector $\vec{q}$. We define the optical conductivity at fintie wavevector $\vec{q}$ as 
\begin{align}
	\vec{j}(\vec{q}, \omega) = \sigma(\vec{q}, \omega) \vec{E}_{\rm ext}(\vec{q}, \omega),
\end{align} 
where $\vec{j}(\vec{q}, \omega)$ is the current density at wavevector $\vec{q}$ and frequency $\omega$. Importantly, we define $\sigma$ as the response to \textit{external} electric field $\vec{E}_{\rm ext}(\vec{q}, \omega)$, which does \textit{not} include the screening field from the induced charges. Then, the quantum weight at finite wavevector $\vec{q}$ is related to the optical conductivity as 
\begin{align}
	\int_0^{\infty}\dd{\omega}\frac{\Re\sigma^{\rm abs}_{ij}(\vec{q}, \omega)}{\omega} &= \frac{\hbar}{2V}K_{ij}(\vec{q}), \label{eq:SWM_q}
\end{align}
where $\sigma^{\rm abs}(\vec{q}, \omega) = (\sigma(\vec{q},\omega) + \sigma(\vec{q},\omega)^{\dagger})/2$ is the absorptive part of the optical conductivity with finite $\vec{q}$. 

We are interested in the limit of $\vec{q}\to 0$. In systems without long-range Coulomb interaction, the optical conductivity $\sigma(\vec{q}, \omega)$ is continuous at $\vec{q}=0$, and therefore we have $\lim_{\vec{q}\to 0}K_{ij}(\vec{q}) = K_{ij}$.

However, in systems with long-range Coulomb interaction, the optical conductivity $\sigma(\vec{q}, \omega)$ at $\vec{q}\to 0$ limit depends on direction of $\vec{q}$ due to the screening effect. For example, in three-dimensional isotropic systems, the longitudinal electric field $\vec{E}_{\rm ext}\parallel \vec{q}$ induces the charge density modulation that screens the electric field, while the transverse electric field $\vec{E}_{\rm ext}\perp \vec{q}$ does not induce such charge modulation. Therefore, the response to the longitudinal and transverse electric fields are different even in the limit of $\vec{q}\to 0$. Writing the optical conductivity as the sum of the longitudinal and transverse components as
\begin{align}
	\sigma_{ij}(\vec{q}, \omega) = \sigma_{L}(\vec{q}, \omega)\frac{q_i q_j}{q^2} + \sigma_{T}(\vec{q}, \omega)\qty(\delta_{ij} - \frac{q_i q_j}{q^2}),
\end{align}
$\sigma_{L}(\vec{q}\to 0, \omega)\neq \sigma_{T}(\vec{q}\to 0, \omega)$ in general. Accordingly, the quantum weight in the limit of $\vec{q}\to 0$ depends on the direction of $\vec{q}$ as well. By decomposing $K_{ij}(\vec{q})$ into the longitudinal and transverse components as
\begin{align}
	K_{ij}(\vec{q}) = K_{L}(\vec{q})\frac{q_i q_j}{q^2} + K_{T}(\vec{q})\qty(\delta_{ij} - \frac{q_i q_j}{q^2}),
\end{align}
we have $K_{L}(\vec{q}\to 0) \neq K_{T}(\vec{q}\to 0)$ in general. 

The static structure factor $S(\vec{q})=(1/V)(\expval{\rho(\vec{q})\rho(-\vec{q})}-\expval{\rho(\vec{q})}\expval{\rho(-\vec{q})})$ is related to the longitudinal part of the optical conductivity~\cite{onishi2025b}. Indeed, noting the continuity equation $\dot{\rho}+\nabla\vdot\vec{j}=0$, and dissipation fluctuation theorem, one can show that the structure factor is related to $K_L$ as 
\begin{align}
	S(\vec{q}) &= \frac{1}{2\pi}\vec{q}^T K_L(\vec{q})\vec{q}. \label{ap:eq:Sq_GL}
\end{align}

Now let us consider the force acting on the probe's capacitor. Here we explicitly consider the position dependence of $\vec{g}_I, \vec{g}$ with wavevector $\vec{q}$ and carefully take the $\vec{q}\to0$ limit. Since $\nabla g_C(\vec{r})$ is always longitudinal, i.e., $\vec{q}\parallel \nabla g_C$ in Fourier space, we have 
\begin{align}
	F &= F_{n_p} + \frac{V}{2\epsilon S} \qty(-\frac{\expval{Q_0^2}}{V} + \frac{e^2}{2\pi} \vec{g}K'(\vec{q}) \vec{g}) \nonumber \\
	&=  F_{n_p} + \frac{V}{2\epsilon S} \qty(-\frac{\expval{Q_0^2}}{V} + \frac{e^2}{2\pi} (\nabla g_C K_L'(\vec{q})\nabla g_C + \vec{g}_I K'(\vec{q}) \vec{g}_I + \nabla g_C K'(\vec{q}) \vec{g}_I + \vec{g}_I K'(\vec{q}) \nabla g_C))  \label{ap:eq:quantum_geometric_force1}
\end{align}
Further rewriting $\expval{Q_0^2}$ with the structure factor by using Eq.~\eqref{ap:eq:Q0_structurefactor} and combining it with Eq.~\eqref{ap:eq:Sq_GL} as well as assuming $\expval{Q_0}\simeq0$, we obtain 
\begin{align}
	F &=  F_{n_p} + \frac{e^2 V}{4\pi \epsilon S} \qty(\vec{g}_I K'(\vec{q}) \vec{g}_I + (\nabla g_C)K'(\vec{q}) \vec{g}_I + \vec{g}_I K'(\vec{q}) \nabla g_C) 
\end{align}
Taking $\vec{q}\to 0$ limit yields the same expression as the results in the main text.

\section{Example: Superconducting resonator as a target}
We can understand the geometric zero-point energy in a language more familiar in superconducting circuits as well. To this end, we replace the electronic system with another superconducting LC circuit. To distinguish the two circuits, we call the original circuit the ``probe'' and the other circuit replacing the electronic system the ``target''. In this case, the coupling through the vector potential is equivalent to an inductive coupling in a circuit language. Indeed, the minimal coupling term $\vec{j}\vdot\hat{\vec{A}}$ becomes $\vec{j}\vdot\vec{g}\hat{\varphi}=(\vec{j}\vdot\vec{g})L_p\hat{I}$ and therefore represents the current-current interaction, which is naturally interpreted as the inductive coupling with mutual inductance $M\sim g L_p$. Therefore, we can analyze the system as two inductively coupled superconducting LC circuits, described by the following Hamiltonian:
\begin{align}
	H_{\rm tot} &= \frac{1}{2}\begin{pmatrix}
		\hat{\Phi} & \hat{\varphi}
	\end{pmatrix}\begin{pmatrix}
		L & M \\
		M & L_p
	\end{pmatrix}^{-1} \begin{pmatrix}
		\hat{\Phi} \\
		\hat{\varphi}
	\end{pmatrix} + \frac{\hat{Q}^2}{2C} + \frac{\hat{q}^2}{2C_p},
\end{align}
where $\hat{\Phi}, \hat{Q}$ are the flux and the charge of the target circuit satisfying $\comm{\hat{\Phi}}{\hat{Q}}=i\hbar$, $L, C$ are the inductance and the capacitance of the target circuit, and $M$ is the mutual inductance between the target and the probe.

Assuming that the frequency of the target is much larger than that of the probe, we can apply our theory to this system. We can rewrite the Hamiltonian as 
\begin{align}
	& H_{\rm tot} = H_{\rm eff}(g\hat{\varphi}) + \frac{\hat{\varphi}^2}{2L_p} + \frac{\hat{q}^2}{2C_p}, \label{eq:Htot2}
\end{align}
where $H_{\rm eff}$ is the effective Hamiltonian for the target in the presence of inductive coupling $g=M/L_p$, given by:
\begin{align}
	H_{\rm eff}(\phi) = \frac{(\hat{\Phi}-\phi)^2}{2L_{\rm eff}} + \frac{\hat{Q}^2}{2C}. \label{eq:Hg_I_SC}
\end{align}
Here $L_{\rm eff} = L - g M$ is the effective inductance of the target. 
The eigenenergy of $H_{\rm eff}(\phi)$ for a given $\phi$ is given by $E_n(\phi) = \hbar\omega_{\rm eff}(n+1/2)$ with $\omega_{\rm eff} = (L_{\rm eff}C)^{-1/2}$. 
$H_{\rm eff}$ in general includes the backaction from the probe to the target through the inductive coupling $g$. When the coupling is sufficiently weak so that $gM = M^2/L_p\ll L$, $L_{\rm eff}\simeq L$ and thus the effect of the probe is only producing an electromagnetic field acting on the target.

With Eq.~\eqref{eq:Hg_I_SC}, we can perform exactly the same analysis as above and we find an additional zero-point energy as 
\begin{align}
    E_{\rm QG} = g^2 \frac{\expval{\hat{Q}^2}}{2C_p} 
    = \frac{g^2}{4} \frac{Z_p}{Z_{\rm eff}} \hbar\omega_p 
\end{align}
where $Z_{\rm eff}=\sqrt{L_{\rm eff}/C}$ and $Z_p=\sqrt{L_p/C_p}$.
In this case, the polarization fluctuation of the target is given by $\expval{\hat{Q}^2}$.

\begin{figure}[htbp]
	\centering
	\includegraphics[width=0.5\linewidth]{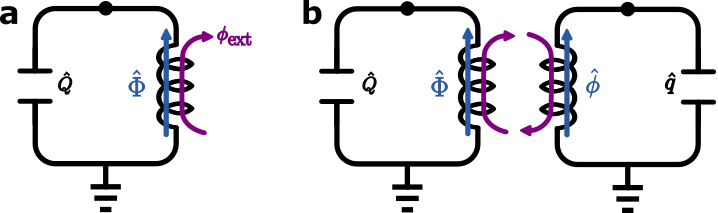}
	\caption{Probing quantum metric with superconducting circuit. (a) A microwave resonator consisting of an inductor and a capacitor with an external magnetic flux $\phi_{\rm ext}$. (b) Two microwave resonators that are inductively coupled.}
	\label{fig:SC_circuits}
\end{figure}